\documentclass[%
 amsmath,amssymb,
 aip,cha,%
reprint,%
]{revtex4-1}

\usepackage{graphicx}
\usepackage{bm}
\usepackage[latin1]{inputenc}
\usepackage{color}
\usepackage{multirow}
\usepackage{capt-of}
\usepackage[caption=false]{subfig}
\makeatletter
\newcommand{\rn}[1]{\romannumeral #1}
\newcommand{\RN}[1]{\expandafter\@slowromancap\romannumeral #1@}
\makeatother
\newcommand{\IN}[1]{~\forall {#1} \in \mathcal{I}_N}
\newcommand{\ING}[1]{~\forall {#1} \in \mathcal{I}_{GL}}
\newcommand{\INC}[1]{~\forall {#1} \in \mathcal{I}_{DL}}

\begin{document}

\preprint{AIP/123-QED}

\title{Modelling, controlling, predicting blackouts}

\author{Chengwei Wang}
 \email{r01cw13@abdn.ac.uk}
 \affiliation{Institute for Complex Systems and Mathematical Biology, University of Aberdeen, King's College, AB24 3UE Aberdeen, United Kingdom.}
\author{Celso Grebogi}
\affiliation{Institute for Complex Systems and Mathematical Biology, University of Aberdeen, King's College, AB24 3UE Aberdeen, United Kingdom.}
\author{Murilo S. Baptista}
\affiliation{Institute for Complex Systems and Mathematical Biology, University of Aberdeen, King's College, AB24 3UE Aberdeen, United Kingdom.}

\date{\today}

\begin{abstract}
The electric power system is one of the cornerstones of modern society.
One of its most serious malfunctions is the blackout, a catastrophic event that may disrupt a substantial portion of the system, playing havoc to human life and causing  great economic losses.
Thus, understanding the mechanisms leading to blackouts and creating a reliable and resilient power grid has been a major issue, attracting the attention of scientists, engineers and stakeholders.
In this paper, we study the blackout problem in power grids by considering a practical phase-oscillator model.
This model allows one to simultaneously consider different types of power sources (e.g., traditional AC power plants and renewable power sources connected by DC/AC inverters) and different types of loads (e.g., consumers connected to distribution networks and consumers directly connected to power plants).
We propose two new control strategies based on our model, one for traditional power grids, and another one for smart grids.
The control strategies show the efficient function of the fast-response energy storage systems in preventing and predicting blackouts in smart grids.
This work provides innovative ideas which help us to build up a robuster and more economic smart power system.
\end{abstract}

\pacs{Valid PACS appear here}
\maketitle

\begin{quotation}
One of the most serious malfunctions of today's electric power grid is the blackout.
A blackout is a phenomenon of cascading failures in power grids that may disrupt a substantial portion of power grids, causing large economic losses and impacting on human life.
Due to the complexity involving in the modelling of the power grid to understand the basic principles leading to blackouts and ways to control it, research on this topic has attracted the attention of not only engineers but also of scientists.
In this paper, we study the blackout phenomena resulting from the synchronisation collapse in the generators, by considering a practical phase-oscillator model, which allows one to simultaneously incorporate different types of power sources and loads.
We propose two smart control strategies, one for traditional power grids in which the control of a generator is solely based on its local state, and another one for smart grids in which a generator is controlled based on information about the state of other relevant components of the grids.
The control strategies aim to show the active influence on the dynamics of smart grids from the fast-response energy storage systems, which provides an innovative approach to mitigate and predict blackouts in smart grids and to build up a robuster and more economic power system.
\end{quotation}

\section{Introduction} 
The electrical infrastructure plays a vital significant role in the modern society.
The blackout, a phenomenon of cascading failures in power grids, is a comprehensive, complicated and fast-evolving process caused by different reasons \cite{pourbeik2006anatomy, nedic2006criticality, andersson2005causes}.
For example, the blackout of the U.S.-Canadian power grid on 14 August, 2003, interrupted approximately 63 GW of load and affected about 50 million people in eight U.S. states and two Canadian provinces \cite{andersson2005causes, dobson2007complex}.
A nationwide blackout happened in Italy on 28 September, 2003, due to cascading failures caused by the tripping of the power transmission line between Italy and Switzerland \cite{bacher2003report, johnson2007analysing}.
On 31 July, 2012, a more severe power blackout caused by a relay problem, affected 22 states of India and left approximate 700 million people in darkness. \cite{lai2012lessons, lai2013investigation}

A great deal of attention from both the engineers \cite{golshani2013implementation, shaobo2010challenges, shortle2014transmission, wang2009cascade, henneaux2013blackout, giraldo2013synchronization} and physicists \cite{menck2014dead, simonsen2008transient, sole2008robustness, albert2004structural, nardelli2014models, rubido2014resiliently} has recently been drawn to study blackouts by considering both traditional and smart grids, aiming at finding the most unifying and fundamental reasons for such events.
Some works \cite{wang2012adaptive, enacheanu2005new, ren2015cascade} proposed advanced control strategies to prevent these events.
Yet, despite these efforts, blackouts are still occurring since power grids are complicated self-organised critical systems \cite{bak1987self, bak1988self, carreras2000initial} experiencing inevitable and diverse levels of disturbances.  
The adverse influence of a blackout tends to increase, since the modern power grids are expanding with more interconnections among different areas and countries \cite{amin2008electric}, making the research for blackout more necessary. 

The collapse of frequency synchronisation (FS) in power systems is one of the main causes behind these catastrophic events.
Collapsing FS is mainly caused by the imbalance of active power between generators and loads \cite{chuvychin2003dynamic, li2012fast}.
In a normal operating state, the active power generation and consumption must be timely equal regardless of the power loss in the system.
Otherwise, some components of the power grid are tripped due to overload resulting in a disconnection between these components and the main network.
A loss of components, such as generators, aggravates the imbalance of active power, which may cause a FS collapse.

In this paper, we discuss the blackout scenarios resulting from an FS collapse, by considering a practical model.
Comparing to the Kuramoto-like model \cite{filatrella2008analysis} and the swing equation \cite{motter2013spontaneous}, our model allows one to study power grids by simultaneously considering different types of power sources and different types of consumers.
We put forward two smart control strategies to avoid it.
One smart control strategy is designed for traditional power grids in which a generator is controlled based on its own state, and another one is for smart grids based on a communication network, which is able to timely collect and exchange information about the state of the network among some important components of power grids.
For the latter control strategy, the behaviour of the controlled power system allows us to predict the power energy that the remaining generators need, to prevent a blackout from happening due to a major failure caused by one generator.
Our control strategies are based on distributed fast-response energy storage systems, which grants a positive motivation for the development of distributed renewable energy. 
Comparing to other works \cite{wang2012adaptive, enacheanu2005new, ren2015cascade}, our control strategies can not only prevent a blackout from happening, but also greatly decrease the requirement of backup power from generators to restore normal functioning of the power systems.
Thus, this work also contributes towards the design and implementation of more resilient and economic power grids.

\section{The Model}
\begin{figure}[h]
\centering
\subfloat[]{\includegraphics[height=2cm]{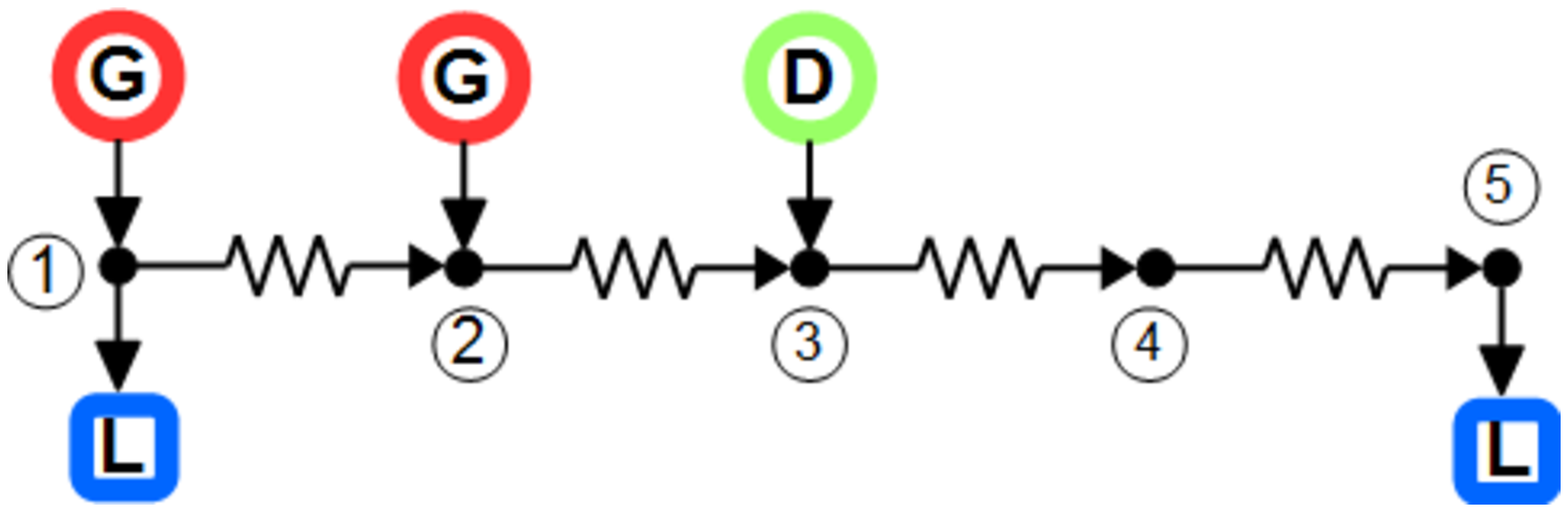}}
\hfill
\subfloat[]{\includegraphics[height=2cm]{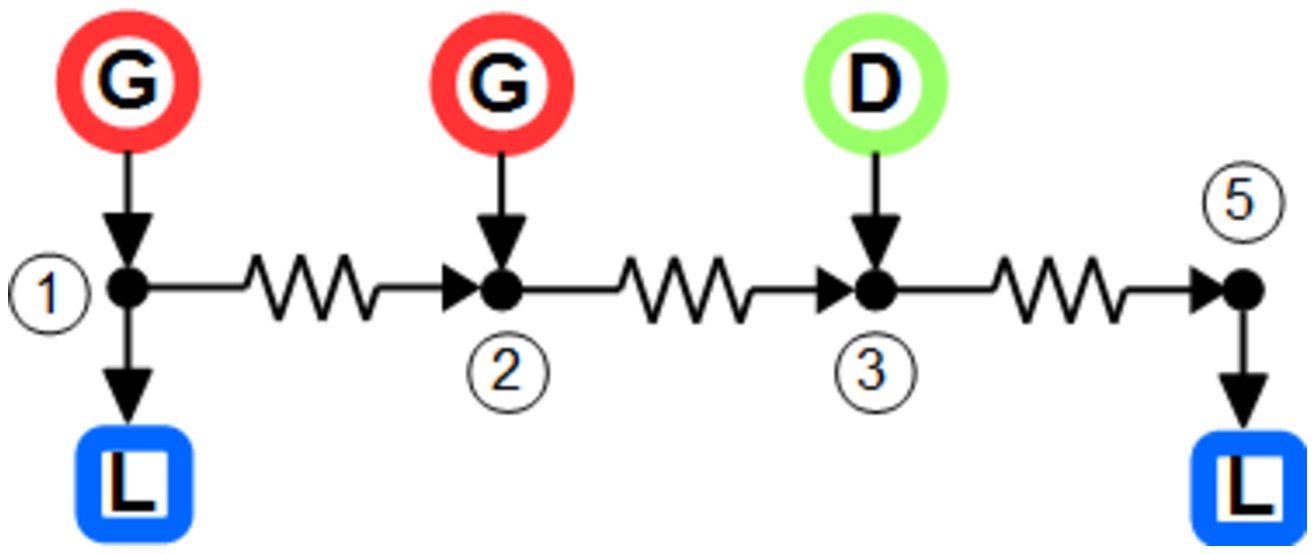}}
\caption{ (a) The original power grid including 2 generators (G), 1 DC source (D), 2 loads (L) and 1 junction node (node 4). (b) The reduced power grid obtained using the Kron reduction to eliminate the junction node.}
\label{fig:simple.power.grid}
\end{figure}

We consider a power grid without power loss in the transmission lines.
A reduced power grid can be obtained by the Kron reduction \cite{caliskan2014towards, caliskan2012kron, dorfler2013kron} that eliminates all of the junction nodes where the input power is equal to the output power [node 4 in Fig.~\ref{fig:simple.power.grid} (a)]. 
Figure~\ref{fig:simple.power.grid} (b) shows the reduced power grid obtained from the one shown in Fig.~\ref{fig:simple.power.grid} (a), using the Kron reduction method to eliminate node 4.
In the reduced network, a load may share a node with a generator [node 1 in Fig.~\ref{fig:simple.power.grid} (b)], or may occupy a separate node [node 5 in Fig.~\ref{fig:simple.power.grid} (b)]. 
We use the elements of the index set $\mathcal{I}_{GL}=\{1,\cdots,N_{GL}\}$ to represent the labels for the nodes indicating generators [node 2 in Fig.~\ref{fig:simple.power.grid} (b)] or the nodes shared by a generator and a load [node 1 in Fig.~\ref{fig:simple.power.grid} (b)], the elements of the index set $\mathcal{I}_{DL}=\{N_{GL}+1,\cdots,N\}$ to denote the labels for the nodes indicating DC sources (e.g., solar power) connected by the DC/AC inverters [node 3 in Fig.~\ref{fig:simple.power.grid} (b)], or the nodes indicating loads occupying separate nodes [node 5 in Fig.~\ref{fig:simple.power.grid} (b)], and the elements of the index set $\mathcal{I}_N=\{1,\cdots,N\}$ to indicate the labels for all nodes in a reduced power grid.

The model of the power grid is given by Eqs.~(\ref{eq:generator}) and (\ref{eq:load}).
The generator is modelled by the swing equation \cite{motter2013spontaneous}, namely,
{\scriptsize
\begin{equation}
\label{eq:generator}
\left\{
\begin{split}
&\dot{\Theta}_i=\Omega_i,\ING{i},\\
&M_i\dot{\Omega}_i=P_{G,i}+P_{L,i}-\sum_{j=1}^N b_{ij}\sin\Theta_{ij}-D_i\Omega_i,\ING{i},\IN{j},
\end{split}
\right.
\end{equation}}where $\Theta_i$ is the phase angle of node $i$, $\Theta_{ij}=\Theta_i-\Theta_j$, 
$\Omega_i$ is the instantaneous angular frequency of generator $i$,
$M_i$ and $D_i$ are the normalised inertia and damping coefficient, respectively, 
$P_{G,i}$ is the mechanical power provided by turbine $i$,
$P_{L,i}$ is the power consumed by the load sharing node $i$ and $P_{L,i}=0$ if there is no load sharing the node with generator $i$,
$b_{ij}=|U_i||U_j|\Im{Y_{ij}}$, 
where $U_i$ is the bus voltage of node $i$, $Y_{ij}=Y_{ji}$ is a complex number representing the admittance of the transmission line between node $i$ and $j$, $Y_{ij}=0$ if $i$ and $j$ are not directly connected in the reduced power grid, and $\Im{Y_{ij}}$ is the imaginary part of $Y_{ij}$. 

(\rn{3}) A DC source or a load occupying a separate node is modelled by \cite{dorfler2013dynamics, giraldo2013synchronization},
\begin{equation}
\label{eq:load}
D_i\dot{\Theta}_i=P_{DL,i}-\sum_{j=1}^N b_{ij}\sin\Theta_{ij},\INC{i},\IN{j},
\end{equation}
where for a DC source, $P_{DL,i}>0$ is the nominal power source, $1/D_i$ is the droop-slope of the droop controller of the DC/AC inverter; for a load occupying a separate node, $P_{DL,i}<0$ is a constant power load, $D_i>0$ and $D_i\dot{\Theta}_i$ is a part of frequency-dependent load.

We apply a rotating frame for the models in Eqs.~(\ref{eq:generator}) and (\ref{eq:load}) by letting $\theta_i=\Theta_i-\Omega_n t$ and $\omega_i=\Omega_i-\Omega_n$, where $\Omega_n=2\pi f_n$ is the natural angular frequency, and $f_n$=50Hz or 60Hz is the natural frequency of the power grid.
The natural angular frequency becomes $\omega_n=0$ in the rotating frame, then Eqs.~(\ref{eq:generator}) and (\ref{eq:load}) become,
{\scriptsize
\begin{equation}
\label{eq:power.grid}
\left\{
\begin{split}
&\dot{\theta}_i=T_{DL,i}-\sum_{j=1}^N a_{ij}\sin \theta_{ij},\INC{i},\IN{j},\\
&\dot{\theta}_i=\omega_i,\ING{i},\\
&\dot{\omega}_i=T_{G,i}+T_{L,i}-\sum_{j=1}^N a_{ij}\sin \theta_{ij}-F_i\omega_i,\ING{i},\IN{j},
\end{split}
\right.
\end{equation}}where $\theta_{ij}=\theta_i-\theta_j$; if $i\in \mathcal{I}_{DL}$, $T_{DL,i}=\frac{P_{DL,i}}{D_i}-\Omega_n$, $a_{ij}=\frac{b_{ij}}{D_i}$; 
if $i\in \mathcal{I}_{GL}$,  $T_{G,i}=\frac{P_{G,i}}{M_i}$, $T_{L,i}=\frac{1}{M_i}(P_{L,i}-D_i\Omega_n)$, $a_{ij}=\frac{b_{ij}}{M_i}$, $F_i=\frac{D_i}{M_i}$. 

Equation~(\ref{eq:power.grid}) describes a power grid of a coupled phase-oscillator network which contains both traditional AC power plants and renewable power sources connected by DC/AC converters, and includes both users connected to distribution networks and consumers powered directly by power plants.
A steady state of the power grid corresponds to a FS state of Eq.~(\ref{eq:power.grid}), defined by $\dot{\theta}_i=\omega_i=0$.  
Summing	 the first and third equations, which are related to power transmission in Eq.~(\ref{eq:power.grid}) for all $i$, we have
{\small
\begin{equation}
\label{eq:sum.all}
\begin{split}
&\sum_{i\in\mathcal{I}_{GL}} \dot{\omega}_i+\sum_{i\in\mathcal{I}_{DL}}\dot{\theta}_i\\
=&\sum_{i\in\mathcal{I}_{GL}}T_{G,i}+\sum_{i\in\mathcal{I}_{GL}} T_{L,i}+\sum_{i\in\mathcal{I}_{DL}} T_{DL,i}-\sum_{i\in\mathcal{I}_{GL}}F_i\omega_i.
\end{split}
\end{equation}}In a steady state, the power system operates at an equilibrium point, and all nodes are in frequency synchronisation with $\dot{\theta}_i^{eq}=\omega_i^{eq}=\omega_n=0$,
implying the ``imbalance power" between generators and loads to be zero, namely, 
\begin{equation}
\label{eq:balance}
d T=\sum_{i\in\mathcal{I}_{GL}}T_{G,i}+\sum_{i\in\mathcal{I}_{GL}} T_{L,i}+\sum_{i\in\mathcal{I}_{DL}} T_{DL,i}=0.
\end{equation}
This means that the power produced by generators is equal to that consumed by loads in a steady state. 

\textit{Blackout process-} Assume that at $t=t_0$ there is a loss of a high-capacity generator with label $m\in\mathcal{I}_{GL}$, i.e., $T_{G,m}$ suddenly becomes $0$ from a large positive value, such that $d T<0$.
The stored kinetic energy in the rotors of all remaining generators are then released to balance the power between generators and consumers, resulting in the deceleration of the speed of rotors, i.e., a drop of the angular frequencies $\omega_i$ from $0$.
In order to maintain the stability of the system, the remaining generators need to provide additional power, such that $dT$ returns to $0$ and all $\omega_i$ also returns to $0$.
The power system then reaches a new steady state.
The regulation process of the output power in a generator can be controlled by its active power regulating system, which is described by,
\begin{equation}
\label{eq:turbine.governor}
\dot{T}_{G,i}=-K_i\omega_i,\ING{i},
\end{equation}
where $K_i>0$ is the regulation constant of generator $i$ that can be manually set,  
and $\omega_i$ indicates the frequency deviation from $0$ for node $i$ in the dynamic process.

Equation (\ref{eq:turbine.governor}) can represent either the turbine governor system or the energy storage system in power grids.
The mechanism of this control is that when $dT<0~(>0)$, the angular speed of the generator rotors, $\omega_i$ decreases (increases) due to the release (accumulation) of kinetic energy in the rotors.
This leads to a negative (positive) $\omega_i$, which, according to Eq.~(\ref{eq:turbine.governor}), forces $T_{G,i}$ to increase (decrease) by automatically turning up (down) the flow rate of steam into the turbine.
Thus, more (less) energy is provided by prime movers.
The adjusting power from prime movers accelerates (decelerates) the rotors to balance the generation power and consumption power.
Finally, $\omega_i$ returns to $0$, $dT$ becomes $0$ as well, and the whole system reaches a new steady state.
By providing such a negative feedback to the system, this control enhances the stability of the system around its equilibrium point.
Traditionally, the turbine governor system needs long time to adjust the flow rate of steam due to the mechanical inertia of machines.
However, some new energy storage systems are developed nowadays \cite{su2011modeling, ribeiro2001energy}, such as large battery arrays, solar farms and the storage systems in wind farms, to provide faster response to the frequency change and quickly provide supplementary power into the system to help it to reach a new steady state.
In this paper, we assume all the energy storage systems in power plants are fast-response systems.

When the system loses generator $m$, the angular frequencies of the remaining generators fluctuate, and these generators provide supplementary power according to Eq.~(\ref{eq:turbine.governor}) in order to mitigate the frequency fluctuation.
Let $T^{max}_{G,i}$ and $T^{min}_{G,i}$ be the upper and lower physical bounds of $T_{G,i}$.
During the dynamic process, if $T_{G,i}>T_{G,i}^{max}$ (overload) or $T_{G,i}<T_{G,i}^{min}$,  generator $i$ is tripped by its protection devices, resulting in a disconnection of an additional generator.
A loss of one more generator results in larger disturbance of the whole system and more power requirement from other generators.
This leads more generators to be tripped due to overload.
Such a cascading failure may lead to a FS collapse, resulting in a blackout in the power grid, i.e., the loss of all generators.
\begin{figure}[h]
\centering
\subfloat{\includegraphics[width=0.32\linewidth]{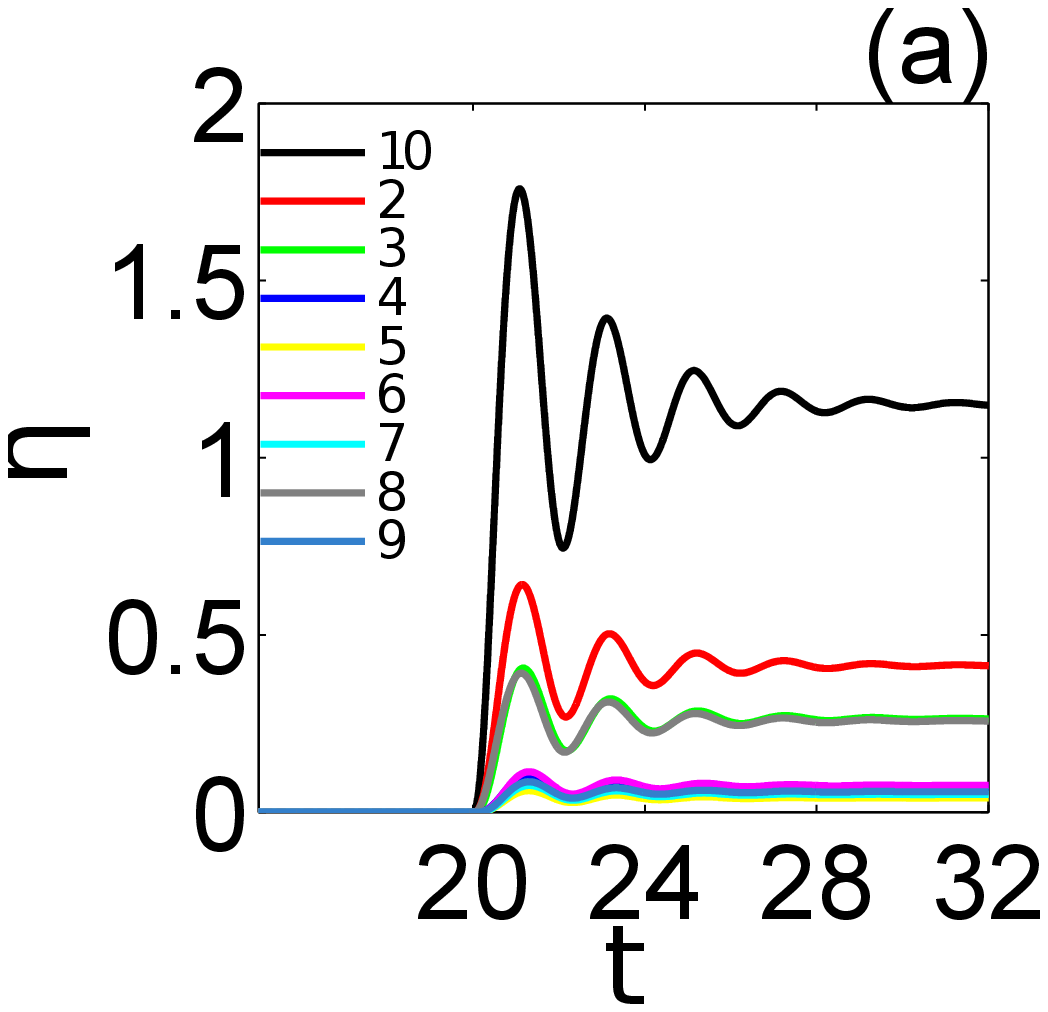}}
\hfill
\subfloat{\includegraphics[width=0.32\linewidth]{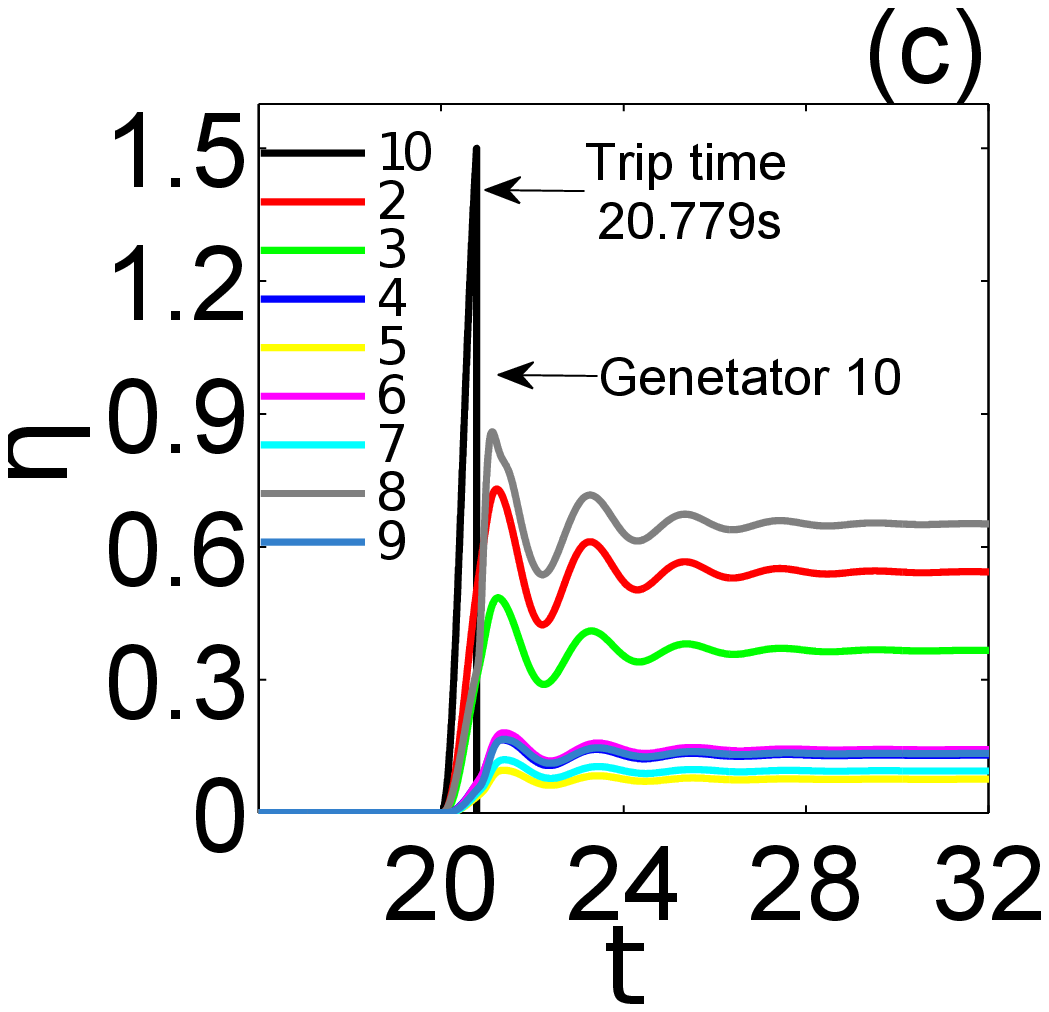}}
\hfill
\subfloat{\includegraphics[width=0.32\linewidth]{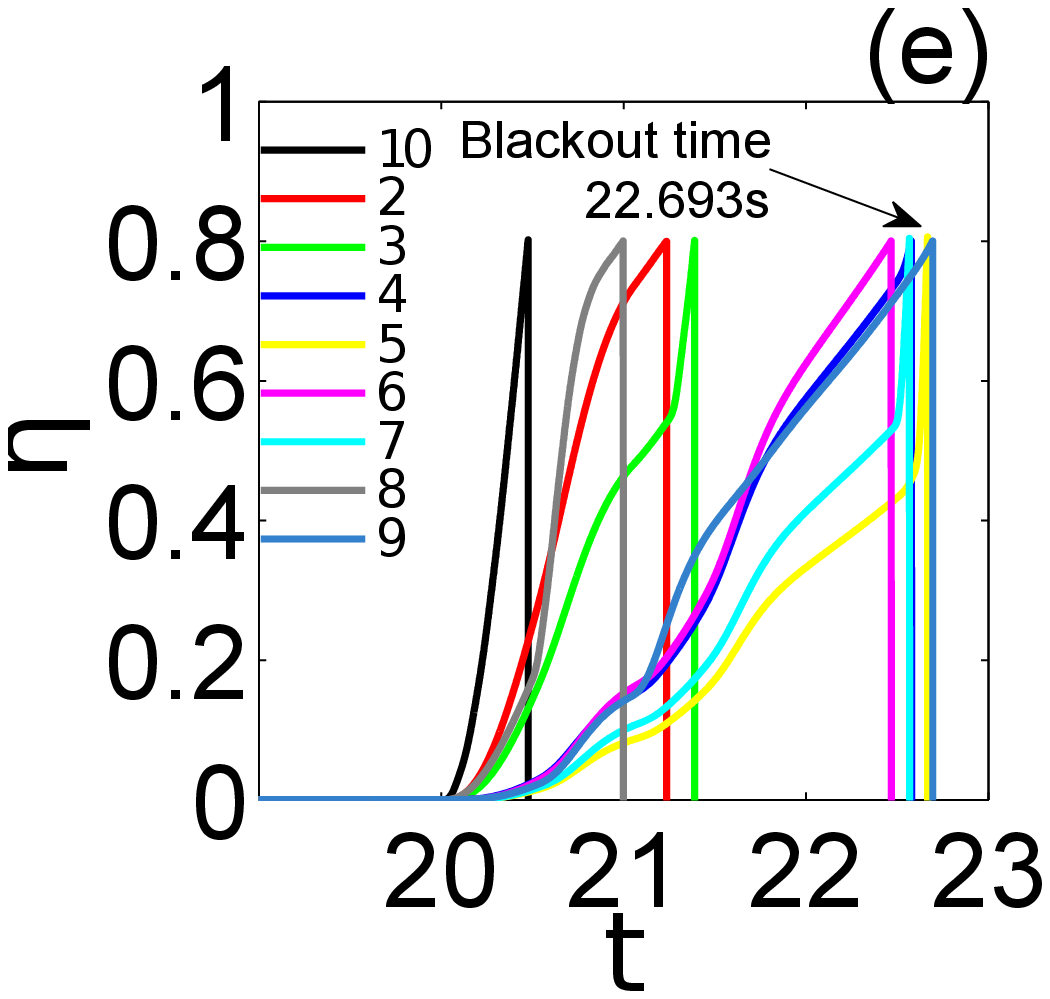}}\\
\subfloat{\includegraphics[width=0.32\linewidth]{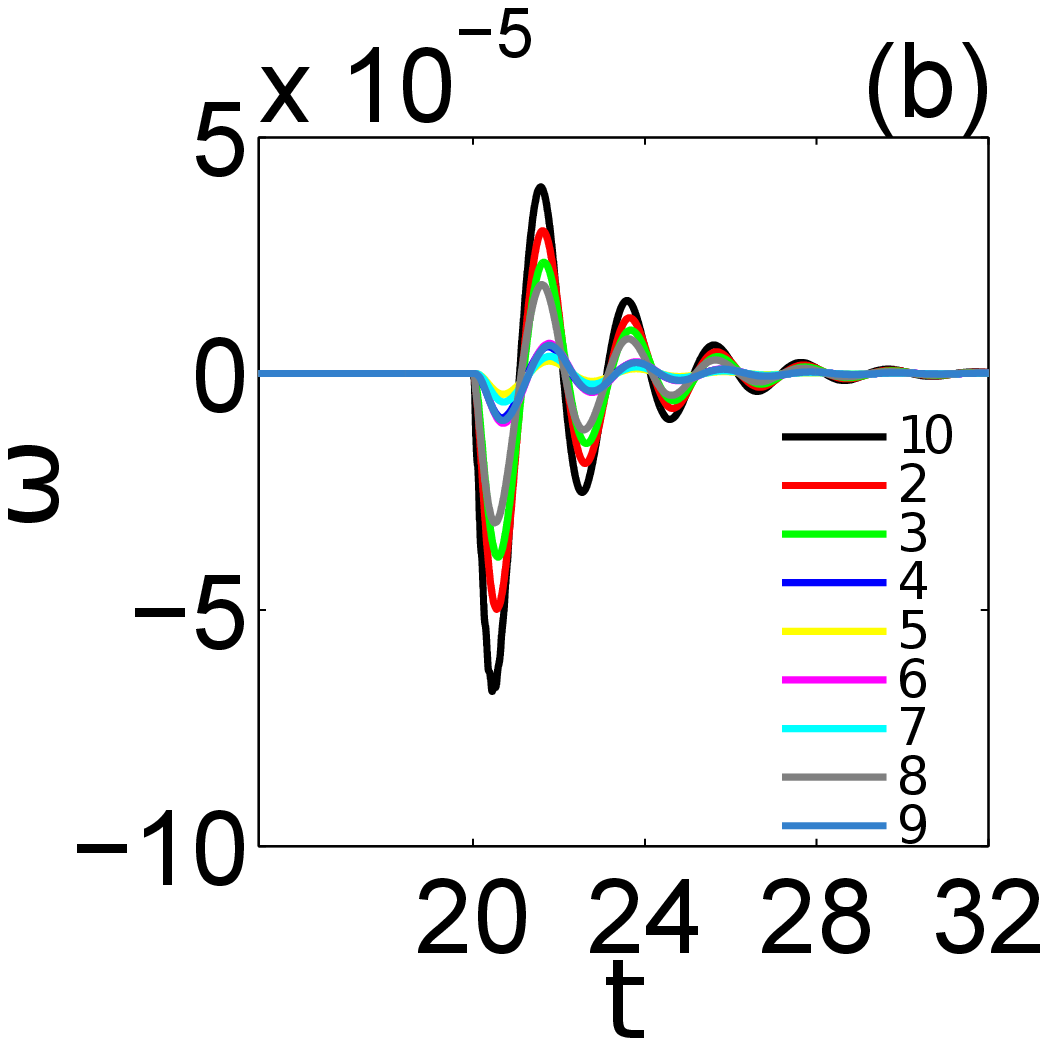}}
\hfill
\subfloat{\includegraphics[width=0.32\linewidth]{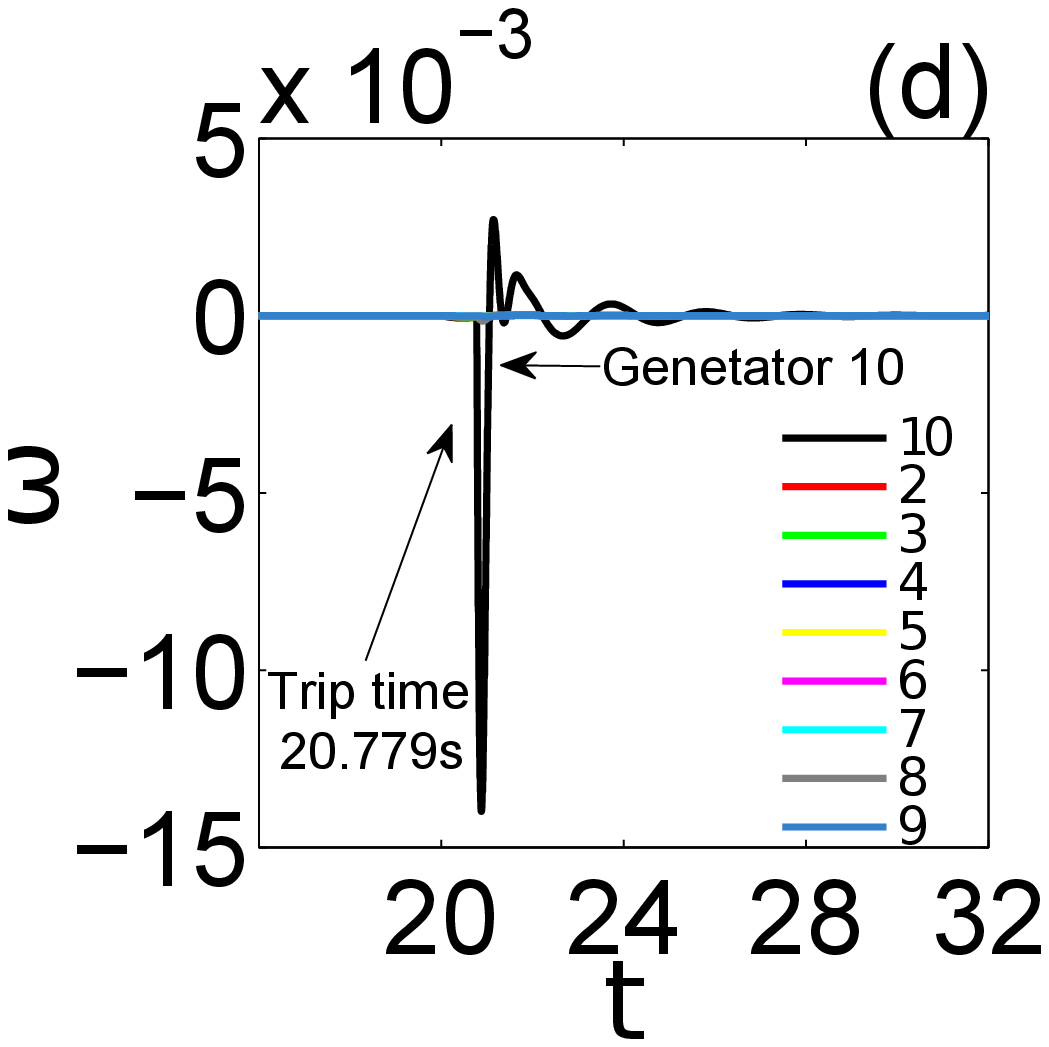}}
\hfill
\subfloat{\includegraphics[width=0.32\linewidth]{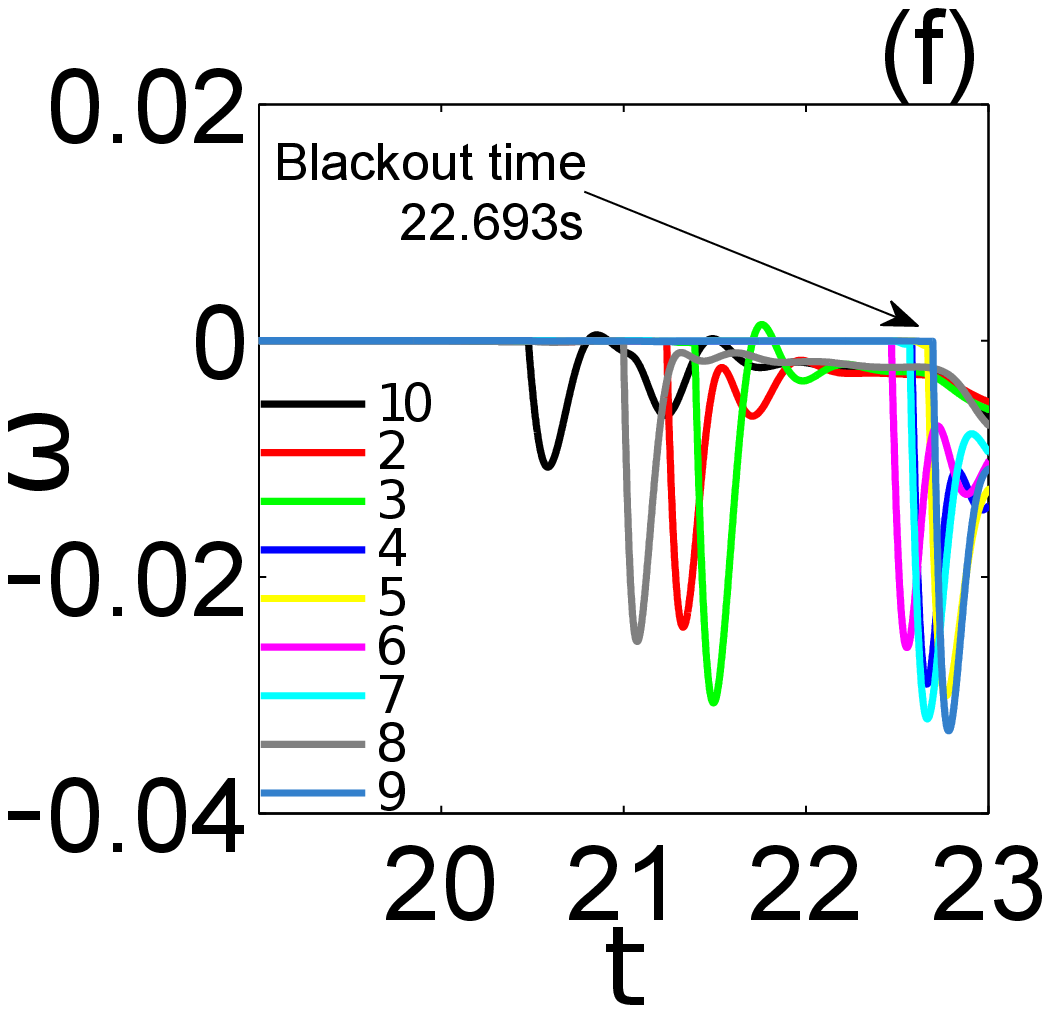}}
\caption{(Color online) Simulation for the IEEE 39 bus system. (a) and (b) show the changes of the power ratio, $\eta_i$, and the angular frequency, $\omega_i$, respectively, with $\eta_i^{max}=2$; (c) and (d) show the changes of $\eta_i$ and $\omega_i$, respectively, with $\eta_i^{max}=1.5$; (e) and (f) show the changes of $\eta_i$ and $\omega_i$, respectively, with $\eta_i^{max}=0.8$.}
\label{fig:simulation}
\end{figure}
The backup-power capacity of generator $i$ is defined by 
\begin{equation}
\label{eq:backuppower}
T_{b,i}=T_{G,i}^{max}-T_{G,i}^{eq}.
\end{equation}
We define the power ratio, $\eta_i$, to indicate the ratio between the power supplemented by generator $i$ and its output power in the steady state before $t_0$ by
\begin{equation}
\label{eq:ratio}
\eta_i=\frac{T_{G,i}-T_{G,i}^{eq}}{T_{G,i}^{eq}}.
\end{equation}
Let 
\begin{equation}
\label{eq:etamax}
\eta_i^{max}=\frac{T_{G,i}^{max}-T_{G,i}^{eq}}{T_{G,i}^{eq}}=\frac{T_{b,i}}{T_{G,i}^{eq}}\geqslant 0
\end{equation}
and 
\begin{equation}
\label{eq:etamix}
\eta_i^{min}=\frac{T_{G,i}^{min}-T_{G,i}^{eq}}{T_{G,i}^{eq}}\leqslant 0
\end{equation}
represent the maximum and minimum values that $\eta_i$ can assume, respectively. 
We simply set $\eta_i^{min}=-0.5$ (i.e, $T_{G,i}^{min}=0.5T_{G,i}^{eq}),~\ING{i}$, because we focus on the overload problem in this paper and $\eta_i^{min}$ does not affect our numerical experiments.
We use the IEEE 39 bus system to show how blackout happens and how $\eta_i^{max}$ affects the behaviour of the system.
The IEEE 39 bus system, also known as the New-England Power System, includes 10 generators, 2 loads sharing nodes with generators, 17 loads occupying separate nodes, and 12 junction nodes.
Appendix A provides the topology and the data required in numerical experiments for this system.
By Kron reduction \cite{caliskan2014towards, caliskan2012kron, dorfler2013kron}, we eliminate the junction nodes and obtain a system with 27 nodes.
In our numerical experiments, we neglect the influence of the voltage change, such that all coupling strengths ($a_{ij}$)	 remain unchanged.
We initiate the dynamic process by switching off generator 1 which has the maximum capacity at t=20s, i.e., forcing $T_{G,1}=0$ at $t=20s$.
We plot the changes of $\eta_i$ and $\omega_i$ for the remaining generators.
We set $T_{G,i}=0$ and $\eta_i=0$ if generator $i$ is tripped due to overload in the experiments.
Figures~\ref{fig:simulation} (a) and \ref{fig:simulation} (b) show the results with a large $\eta_i^{max}=2.0,~\ING{i}$. 
Every generator supplements some power from $t=20s$ and no remaining ones are tripped [Fig.~\ref{fig:simulation} (a)]. 
The angular frequency of each generator, $\omega_i$, experiences fluctuation from $t=20s$, but finally returns to $0$ [Fig.~\ref{fig:simulation} (b)], meaning that the system reaches a new steady state.
Figures~\ref{fig:simulation} (c) and \ref{fig:simulation} (d) show simulations considering $\eta_i^{max}=1.5,~\ING{i}$.
Figure~\ref{fig:simulation} (c) indicates that generator 10 is tripped at $t=20.779s$ due to overload, but other generators successfully provides enough power to the system.
Thus, the system reaches a new steady state, i.e., the angular frequencies of the remaining generators finally become $0$, as shown in Fig.~\ref{fig:simulation} (b).
Figures~\ref{fig:simulation} (e) and \ref{fig:simulation} (f) show the result with $\eta_i^{max}=0.8,~\ING{i}$. 
As shown in Fig.~\ref{fig:simulation} (e), the generators are tripped one by one due to overload.
Finally, at $t=22.693s$, the system lose all generators and a FS collapse occurs [Fig.~\ref{fig:simulation} (f)]. 

\section{Smart control}
\subsection{Smart control \RN{1}}
As shown in Figs.~\ref{fig:simulation} (c) and \ref{fig:simulation} (e), when a generator is tripped due to overload, the $\eta_i$ of the remaining generators are still far away from the maximum limit, meaning that the remaining generators still possesses large amounts of backup power that can be used to restore the stability of the power grid.
In order to efficiently use the backup power of every generator to avoid a blackout, we develop a smart control, which will greatly improve the robustness of power grids with  less requirement of backup power for generators. 
For that, we	 change Eq.~(\ref{eq:turbine.governor}) into 
\begin{equation}
\label{eq:smart.control}
\dot{T}_{G,i}=-\alpha_i K_i\omega_i,~\ING{i},
\end{equation}
where $\alpha_i=(T_{G,i}^{max}-T_{G,i})/(T_{G,i}^{max}-T_{G,i}^{eq})$ if $\omega_i\leqslant 0$, and $\alpha_i=(T_{G,i}-T_{G,i}^{min})/(T_{G,i}^{eq}-T_{G,i}^{min})$ if $\omega_i>0$.
At a steady state, $T_{G,i}=T_{G,i}^{eq}$, we have $\alpha=1$; when generator $i$ reaches its output limits, we have $T_{G,i}=T_{G,i}^{max}$ or $T_{G,i}=T_{G,i}^{min}$ resulting in $\alpha_i=0$.
Thus, $T_{G,i}$ does not change any more when it reaches its limits, and none of the generators are then tripped due to overload.
The utilisation ratio of the backup-power capacity of generator $i$ is defined by 
\begin{equation}
\label{eq:utilisation}
\sigma_i\%=\frac{T_{G,i}-T_{G,i}^{eq}}{T_{b,i}}
\end{equation}
Figure~\ref{fig:smart1} and Tab.~\ref{table1} show the numerical results for the IEEE 39 bus system with smart control \RN{1}.
Set $\eta_i^{max}=0.2,~\ING{i}$, which is smaller than $\eta_i^{max}=0.8,~\ING{i}$ that was used in Figs.~\ref{fig:simulation} (e) and \ref{fig:simulation} (f) where a blackout happens without the implementation of 	smart control \RN{1}. 
A smaller $\eta_i^{max}$ indicates a smaller backup-power capacity of generator $i$.
At $t=20s$ we lose generator 1.
With our control strategy, none of the remaining generators is tripped, although some of them have almost provided their full backup-power capacity (some $\sigma_i\% \approx 100\%$ in Tab.~\ref{table1}),
and the angular frequencies of all remaining generators return to $0$ after some fluctuations [Fig.~\ref{fig:smart1} (b)].
This means that, by applying the smart control \RN{1}, we avoid a blackout in the system with less backup-power capacity requirement for generators.
Less backup-power capacity requirement greatly improves the economic side of power systems.
\\ 

\noindent\begin{minipage}{\linewidth}
\noindent\begin{minipage}[b]{\linewidth}
\centering
\includegraphics[width=0.48\linewidth]{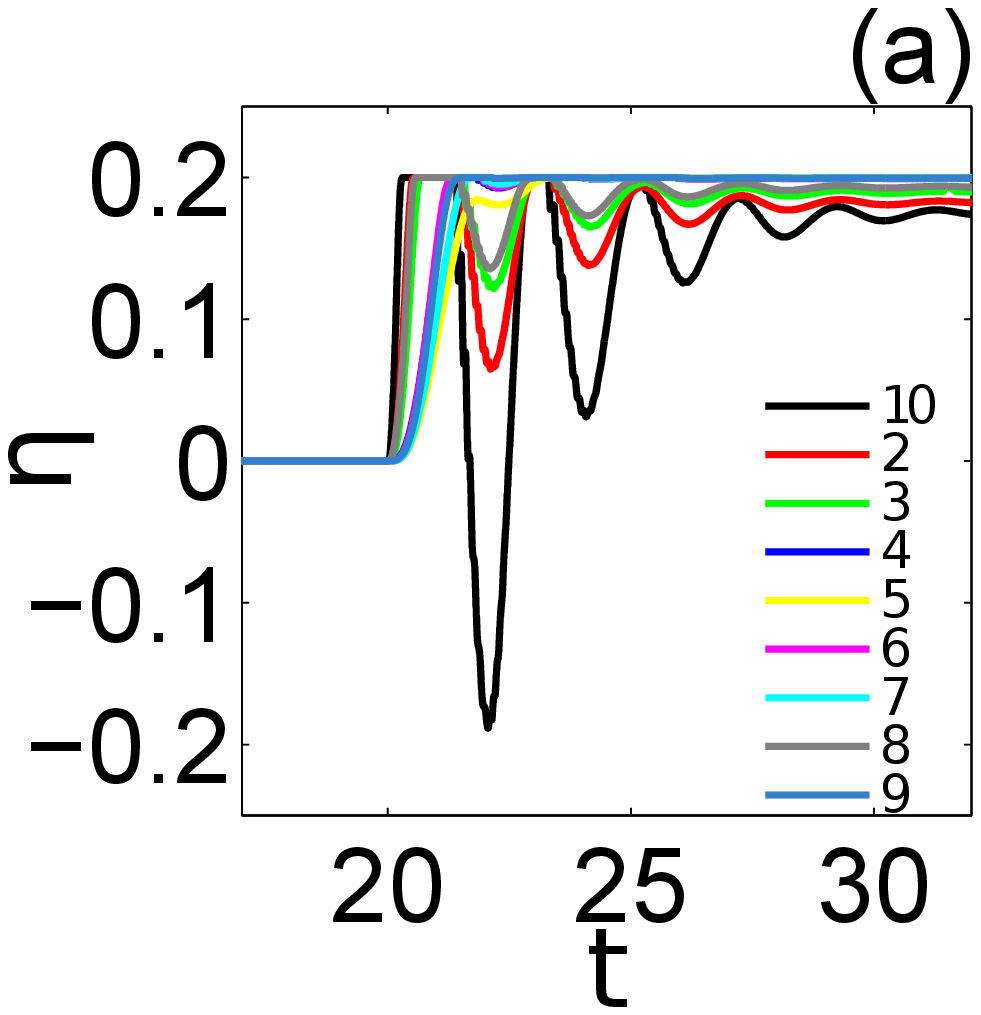}
\includegraphics[width=0.48\linewidth]{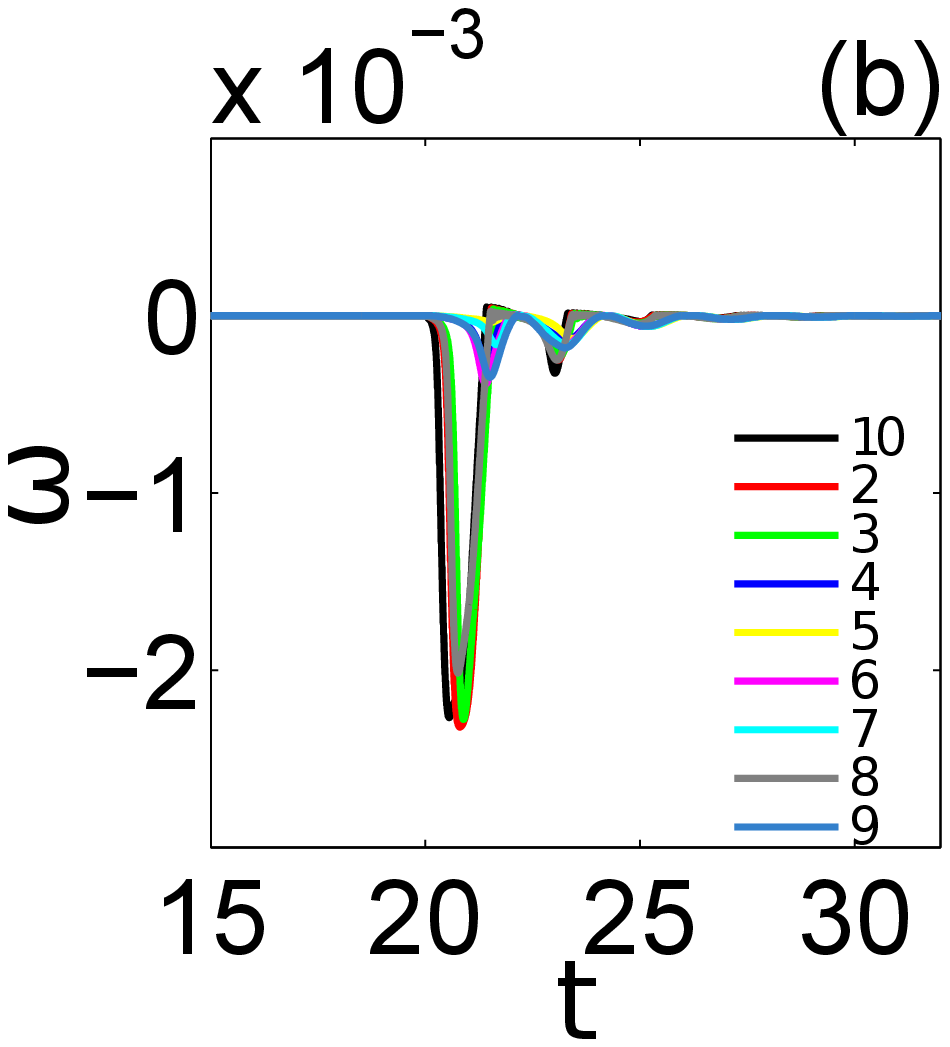}
\captionof{figure}{(Color online) Results for the IEEE 39 bus system with smart control \RN{1}. (a) and (b) show the changes of the power ratio, $\eta_i$, and the angular frequency, $\omega_i$, respectively, with $\eta_i^{max}=0.2$.}
\label{fig:smart1}
\end{minipage}\\
\noindent\begin{minipage}[b]{\linewidth}
\centering
\captionof{table}{Utilisation ratio of backup power of generators with smart control \RN{1}}    
\label{table1}
\begin{tabular}{|l|c|c|c|c|c|c|c|c|c|}
\hline
$i$          & 10   & 2    & 3    & 4    & 5    & 6    & 7    & 8    & 9    \\ \hline
$\sigma_i\%$ & 87.9 & 91.4 & 94.9 & 99.7 & 99.8 & 99.6 & 99.8 & 96.7 & 99.7 \\ \hline
\end{tabular}
\end{minipage}
\end{minipage}\\

\subsection{Smart control \RN{2}} The smart control \RN{1} is easily implemented because $\omega_i$ and $\alpha_i$ can be locally measured or calculated for every generator. 
However, the drawback of the smart control \RN{1} is that some of the generators nearly reach their maximum output limits (some $\sigma_i\%\approx 100\%$ in Tab.~\ref{table1}), reserving no extra power for engineers to impose any further manual control with these generators and leaving these generators at a dangerous critical state.
Next, we propose an improved smart control based on smart grid technology to tackle this problem.
With the fast development of smart grids, it becomes possible to timely measure and exchange the information (e.g., $\omega_i$) among different nodes in a power network by a separate network layer -- the communication network \cite{giraldo2013synchronization, kuhnlenz2015dynamics}. 
To utilise the communication network, we change Eq.~(\ref{eq:turbine.governor}) into
\begin{equation}
\label{eq:smart.control2}
\dot{T}_{G,i}=-\beta_i K_i \overline{\omega},
\end{equation}
where $\beta_i=T_{b,i}/T_b^{max}$ with $T_b^{max}:=\max\{T_{b,i}|i\in\mathcal{I}_{GL}\}$ indicating the maximum  backup-power capacity among all generators, and $\overline{\omega}=\sum_{i=1}^N \gamma_i\omega_i/\sum_{i=1}^N \gamma_i$ with $\gamma_i$ indicating the importance level of node $i$.
We set $\gamma_i= 1$ if node $i$ is a generator, a large capacity DC source or an important load which is sensitive to frequency change, and $\gamma_i=0$ if the information of node $i$ is unavailable or node $i$ is not important (e.g., a normal load).

Smart control \RN{2} improves the control performance by introducing the  average angular frequency, $\overline{\omega}$, which embodies a teamwork principle, i.e., one generator lost, all the remaining generators supplement required power together, according to the change of the average angular frequency of some important nodes instead of according to their local angular frequencies.
Furthermore, the new variable, $\beta_i$, prompts generator $i$ to provide power based on its backup-power capacity, $T_{b,i}$.
In other words, a generator with a larger backup-power capacity contributes more power to the power grid than the one with a smaller capacity. 
This is also an improvement compared to the smart control \RN{1} in which the variable $\alpha_i$ just limits the maximum output of generator $i$ to ensure its non-overload.
\\

\noindent\begin{minipage}[t]{\linewidth}
\noindent\begin{minipage}[t]{\linewidth}
\centering
\includegraphics[width=0.48\linewidth]{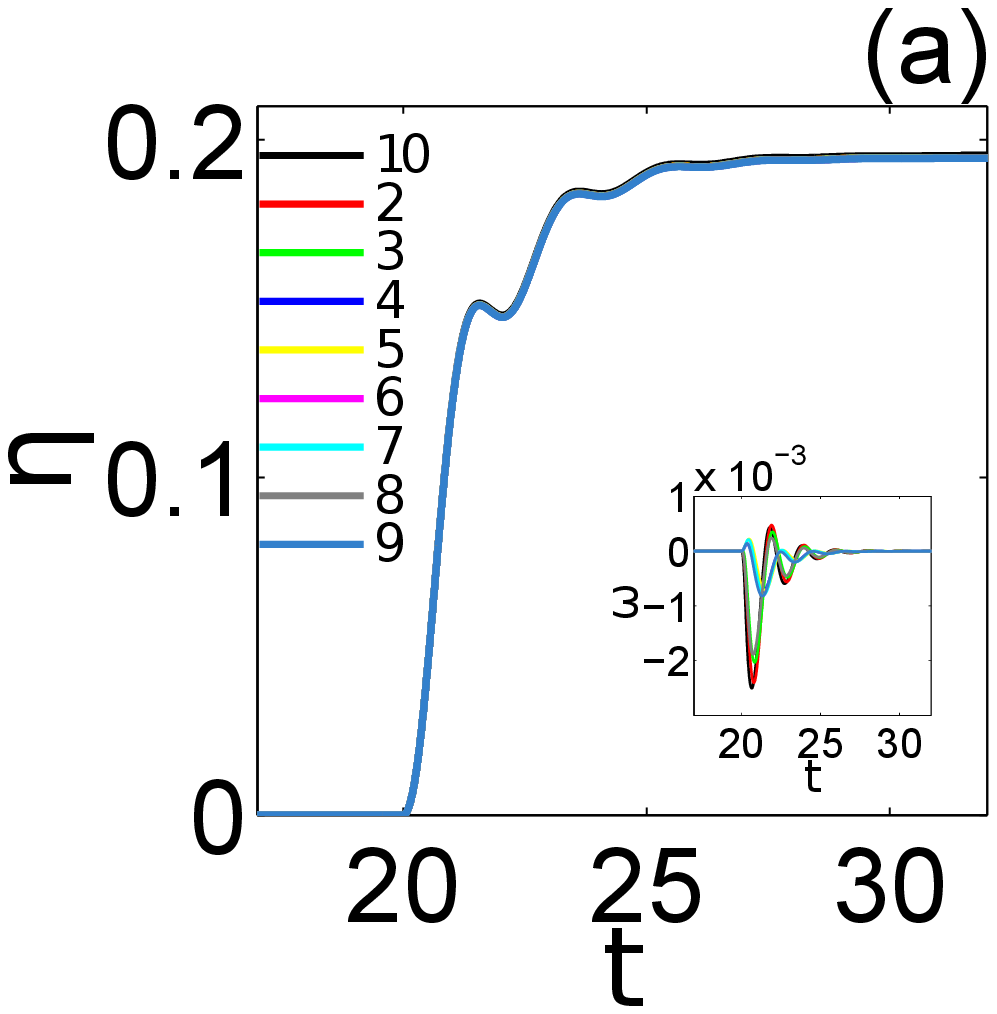}
\includegraphics[width=0.48\linewidth]{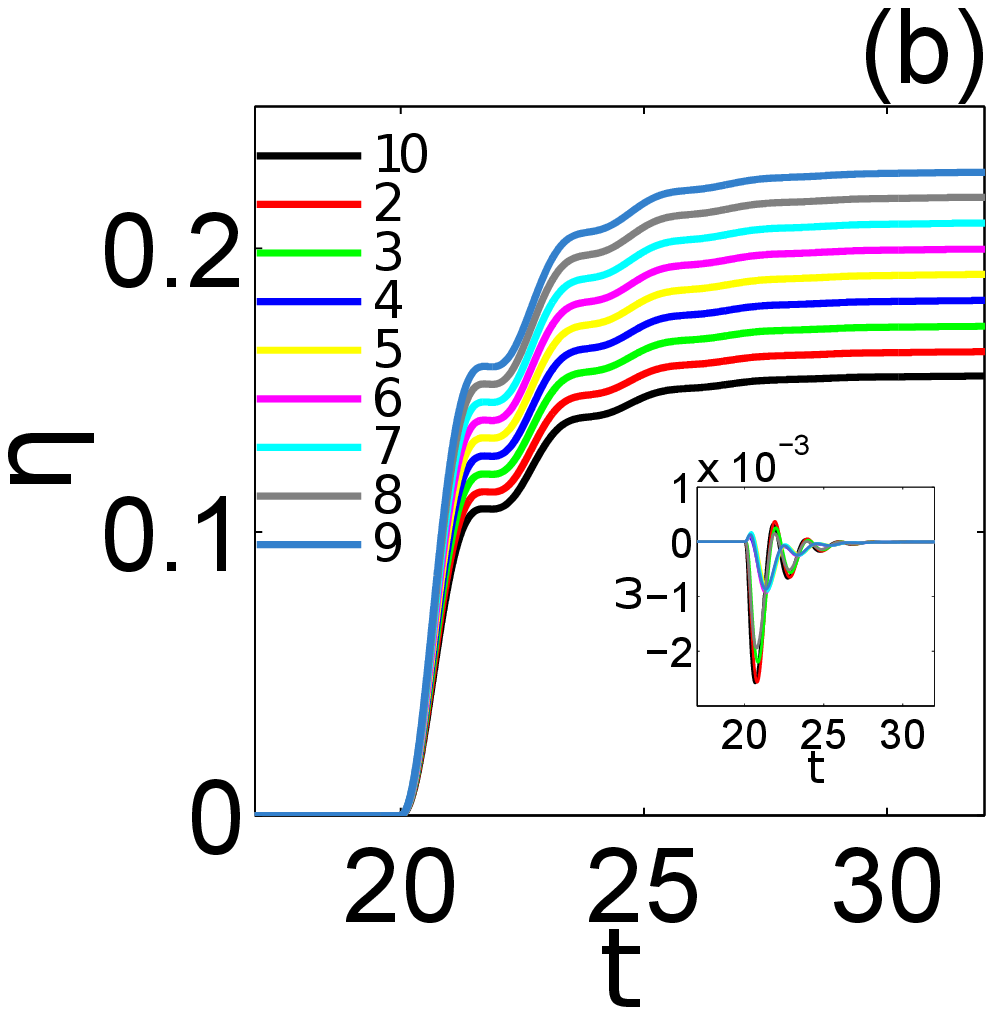}
\captionof{figure}{(Color online) Simulation for the IEEE 39 bus system with the smart control \RN{2}. (a) shows the change of the power ratio, $\eta_i$, by the main plot and the angular frequency, $\omega_i$ by the sub-plot with $\eta_i^{max}=0.2,~\ING{i}$; (b) shows the changes of $\eta_i$ (main plot) and $\omega_i$ (sub-plot) with $\eta_i^{max}$ varying from 0.17 to 0.25.}
\label{fig:smart2}
\end{minipage}\\
\noindent\begin{minipage}[t]{\linewidth}
\centering
\captionof{table}{Utilisation ratio of backup power of generators with smart control \RN{2}}    
\label{table2}
\begin{tabular}{|l|c|c|c|c|c|c|c|c|c|}
\hline
$i$            & 10   & 2    & 3    & 4    & 5    & 6    & 7    & 8    & 9    \\ \hline
$\eta_i^{max}$ & \multicolumn{9}{c|}{0.20}                                    \\ \hline
$\sigma_i\%$   & 97.7 & 97.4 & 97.3 & 97.3 & 97.4 & 97.3 & 97.4 & 97.4 & 97.3 \\ \hline\hline
$\eta_i^{max}$ & 0.17 & 0.18 & 0.19 & 0.20 & 0.21 & 0.22 & 0.23 & 0.24 & 0.25 \\ \hline
$\sigma_i\%$   & 91.2 & 90.9 & 90.8 & 90.8 & 90.9 & 90.8 & 90.9 & 90.9 & 90.8 \\ \hline
\end{tabular}
\end{minipage}
\end{minipage}\\

Figure \ref{fig:smart2} and Tab.~\ref{table2} demonstrates the effectiveness of smart control \RN{2}. 
At $t=20s$, we lose generator 1.
Figure~\ref{fig:smart2} (a) shows the results with $\eta_i^{max}=0.2,~\ING{i}$.
The nine lines indicating $\eta_i$ for the nine remaining generators in Fig.~\ref{fig:smart2} (a) merges into one and $\sigma_i\%\approx 97\%$ for $i=2,\cdots,10$, as shown in Tab.~\ref{table2}, meaning that all generators supplement power with the same ratio to their back-up power capacities.
Finally, none of the remaining generators reaches their full output limits, and the frequencies finally return to $0$ as shown in the sub-plot in Fig.~\ref{fig:smart2} (a). 
In real power systems, $\eta_i^{max}$ is not strictly equal to $\eta_j^{max}$ for $i\neq j$.
Figure~\ref{fig:smart2} (b) shows the result for a more realistic case where $\eta_i^{max}$ varies from 0.17 to 0.25 (shown in Tab. \ref{table2}). 
In this case, $\sigma_i\%\approx91\%$ for $i=2,\cdots,10$ (shown in Tab.~\ref{table2}), meaning that all remaining generators still provide power with the same proportion to their own backup-power capacities, even though $\eta_i^{max}$ is different.  
Our numerical experiments indicate that smart control \RN{2} not only avoids a blackout, but also prevents some generators from reaching their critical points, which greatly improves the stability and robustness of the IEEE 39 bus system.

By comparing Fig.~\ref{fig:simulation} (a) and Fig.~\ref{fig:smart2}, we conclude that smart control \RN{2} (FIG.~4) restrains oscillations on the curves of $\eta$.
This means that, with the implementation of the smart control \RN{2}, the remaining generators do not need to provide more power in the dynamic process than that required in the final steady state of the power system. 
Thus, the back-up power capacity of generators can be decreased by implementing the smart control \RN{2}, which greatly increases the economic side of power systems.

\section{Predicting blackouts}
Assume that the generator $m$ with capacity $T_m$ is lost.
Smart control \RN{1} and \RN{2} enable the remaining generators to provide their full back-up power capacities.
As a consequence, we can predict that a blackout happens if the total back-up power $\sum_{i\neq m} T_{b,i}$ cannot match the lost capacity, i.e., if $\sum_{i\neq m} T_{b,i}<T_m $.
Furthermore, smart control \RN{2} allows the remaining generators to provide power with nearly the same ratio ($\sigma\%=\sigma_i\%\approx \sigma_j\%$, for $i\neq j$) as their back-up power capacities, $T_{b,i}$.
This means that $T_{m}=\sum_{i\neq m}\sigma_i\% T_{b,i}\approx \sigma\%\sum_{i\neq m} T_{b,i}$, i.e, the utilisation ratio of the back-up power capacity for every remaining generator can be approximately obtained by
\begin{equation}
\sigma\%\approx\frac{T_{m}}{\sum_{i\neq m} T_{b,i}}.
\end{equation} 
Thus, we can predict, without numerical simulation, how much power is finally provided by each remaining generator by changing Eq.~(\ref{eq:utilisation}) to
\begin{equation}
\label{eq:predict}
T_{G,i}=\sigma\% T_{b,i}+T_{G,i}^{eq},
\end{equation}
where $T_{b,i}$ can be calculated from Eq.~(\ref{eq:etamax}) with known $\eta_i^{max}$ and $T_{G,i}^{eq}$.

Define the relative error between the predicted and the numerical obtained values of $T_{G,i}$ by $\delta_i=|(T_{G,i}'-T_{G,i})/T_{G,i}|$, where $T_{G,i}'$ indicates the predicted value of $T_{G,i}$ from Eq.~(\ref{eq:predict}).
We carry out two numerical simulations similar to the previous ones.
We set $\eta_i^{max}=0.2,~\ING{i}$ for one case, and set $\eta_i^{max}$ varying from 0.17 to 0.25 for another case.
At $t=20s$, we lose generator $1$.
We record the numerical output power of the remaining generators after the system is restored.
Table~\ref{table3} demonstrates the values of $\delta_i$ in the simulations.
All the values of $\delta_i$ are small, which means that our prediction is effective.
\begin{minipage}[t]{\linewidth}
\centering
\captionof{table}{The relative errors between the predicted and numerical values of $T_{G,i}$ with smart control \RN{2}}    
\label{table3}
\begin{tabular}{|l|c|c|c|c|c|c|c|c|c|}
\hline
$i$            & 10   & 2    & 3    & 4    & 5    & 6    & 7    & 8    & 9    \\ \hline
$\eta_i^{max}$ & \multicolumn{9}{c|}{0.20}                                    \\ \hline
$\delta_i/10^{-4}$   & 5.4 & 0.3 & 1.3 & 1.3 & 0.3 & 1.3 & 0.3 & 0.3 & 1.3 \\ \hline\hline
$\eta_i^{max}$ & 0.17 & 0.18 & 0.19 & 0.20 & 0.21 & 0.22 & 0.23 & 0.24 & 0.25 \\ \hline
$\delta_i/10^{-4}$   & 4.8 & 0.4 & 1.1 & 1.2 & 0.5 & 1.3 & 0.5 & 0.5 & 1.5 \\ \hline
\end{tabular}
\end{minipage}

\section{Conclusion}
In this paper, we have studied the mechanism that creates blackouts in a realistic model for the power grid due to a loss of synchronisation among the generators.
Based on this study, we provided two smart control strategies which requires less backup power for the generators to avoid the onset of a blackout.
One of the smart control strategies was used for the traditional power systems, in which the control of a generator is only based on its own state; the other control was designed for the smart grids, in which the control of a generator considers the state of other generators. 
For the latter control strategy, the behaviour of the controlled power system allowed us to predict the power energy that the remaining generators needed, to prevent a blackout from happening due to a major failure caused by one generator. 
Our control strategies demonstrate the active influence of the distributed fast-response storage systems in smart grids.

We considered the IEEE 39 bus system as a practical topology for simulations, instead of an abstract topological model of the power grids, such as small world networks or random networks.
However, our control methods were applied to the fast-response energy storage systems in power plants, regardless of the topology of the network.
Thus we can safely conclude that they are robust for  power grids with arbitrary topology.
Our work contributes for the understanding of power grids by studying a more practical model, also helps engineers to improve the robustness and economic aspect of power grids.

\begin{acknowledgments}
C.-W.W. is supported by a studentship funded by the College of Physical Sciences, University of Aberdeen. M.S.B. acknowledges EPSRC grant NO. EP/I032606/1.
\end{acknowledgments}

\appendix

\section{Appendixes}
\begin{figure}[h]
\centering
\includegraphics[width=\linewidth]{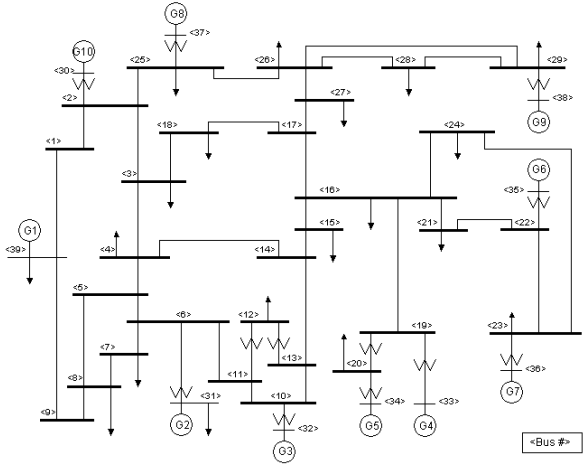}
\caption{Diagram of the IEEE 39 bus system. Image obtained from Ref. \cite{IEEE39}.}
\label{fig:IEEE39}
\end{figure}

The topology of the IEEE 39 bus system is shown by Fig.~\ref{fig:IEEE39}. 
We treat every bus as a node, thus, there are 39 nodes in this system including 10 generators, 17 consumers occupying separated nodes, 2 consumers sharing nodes with generators (bus 31 and 39), 12 junction nodes.
This power grid can be reduced to a 27 node network by eliminating the 12 junction nodes through Kron reduction.
In our analysis, we use $N^0=39$ and $N=27$ to represent the total number of nodes in the original network and in the reduced network, respectively.

Tables~\ref{table4} and \ref{table5} give the data for buses and transmission lines, respectively.
Reference \cite{IEEE39} provides a power flow study result for the IEEE 39 bus system, which contains all the original data, except for the damping coefficient $D$ and the control parameter $K$, for our numerical experiments. 
We set $D_i=2,~\ING{i}$, $D_i=3,~\INC{i}$,
Let $K_i=1,~\ING{i}$ for the numerical experiments without smart control strategy (SCS) and with SCS \RN{1},
and $K_i=0.01,~\ING{i}$ for the numerical experiments with SCS \RN{2}, since generators become more sensitive to the frequency change with SCS \RN{2} and the numerical experiments cannot get a convergent result with $K=1$.
The normalised inertia can be calculated by $M_i=\frac{2H_i}{\Omega_n}=\frac{2H_i}{2\pi f_n}$, where $f_n=$50Hz or 60Hz is the natural frequency of power systems. 
$P_G^0$ and $P_L^0$ are obtained from the power flow results in Ref.~\cite{IEEE39} by considering with power loss in transmission lines, resulting in $\sum_{i=1}^{N_G} P_{G,i}^0+\sum_{i=N_G+1}^{N} P_{L,i}^0=43.71\neq 0$, where 43.71 indicates the total power loss in transmission lines.
In order to construct a reduced network (27 nodes) without power loss in transmission line, we let $P_{G,i}=P_{G,i}^0-\overline{P}$ and $P_{L,i}=P_{L,i}^0-\overline{P}$, where $\overline{P}=\frac{1}{N}\left(\sum_{i=1}^{N_G}P_{G,i}^0+\sum_{i=N_G+1}^{N} P_{L,i}^0\right)=\frac{43.17}{27}\approx 1.6189$.
This means that the 27 nodes in the reduced network share equally the power loss in transmission lines, such that $\sum_{i=1}^{N_G}P_{G,i}+\sum_{i=N_G+1}^{N} P_{L,i}=0$,  indicating that the power provided by generators are equal to that consumed by consumers.

We neglect the reactances of all generators and the transformer tapping when we calculate the coupling strengths, although they are provided in Ref.~\cite{IEEE39}, because this neglect dose not affect the results of our numerical experiments, but  greatly simplifies the experiments.   
Thus, the coupling strengths can be calculated by the following steps: (\rn{1}), using the data in Tab.~\ref{table4} to calculate the admittance matrix $\mathbf{Y}$, where $Y_{pq}=Y_{qp}=-\frac{1}{R_{pq}+iX_{qp}}$ if $p\neq q$, and $Y_{pp}=-\sum_{q=1}^{N^0} Y_{pq}+\frac{1}{2}iB_{pq}$; (\rn{2}) using $U$ in Tab.~\ref{table5} to calculate $b_{pq}$ by $b_{pq}=|U_p||U_q|\Im Y_{pq}$ where $\Im Y_{pq}$ is the imaginary part of $Y_{pq}$.

\begin{table}[]
\centering
\caption{Transmission line data. Data from Ref. \cite{IEEE39}.}
\label{table4}
\begin{tabular}{|c|c|c|c|c|}
\hline
From Bus & To Bus & R      & X      & B      \\ \hline
1        & 2      & 0.0035 & 0.0411 & 0.6987 \\ \hline
1        & 39     & 0.0010 & 0.0250 & 0.7500 \\ \hline
2        & 3      & 0.0013 & 0.0151 & 0.2572 \\ \hline
2        & 25     & 0.0070 & 0.0086 & 0.1460 \\ \hline
3        & 4      & 0.0013 & 0.0213 & 0.2214 \\ \hline
3        & 18     & 0.0011 & 0.0133 & 0.2138 \\ \hline
4        & 5      & 0.0008 & 0.0128 & 0.1342 \\ \hline
4        & 14     & 0.0008 & 0.0129 & 0.1382 \\ \hline
5        & 6      & 0.0002 & 0.0026 & 0.0434 \\ \hline
5        & 8      & 0.0008 & 0.0112 & 0.1476 \\ \hline
6        & 7      & 0.0006 & 0.0092 & 0.1130 \\ \hline
6        & 11     & 0.0007 & 0.0082 & 0.1389 \\ \hline
7        & 8      & 0.0004 & 0.0046 & 0.0780 \\ \hline
8        & 9      & 0.0023 & 0.0363 & 0.3804 \\ \hline
9        & 39     & 0.0010 & 0.0250 & 1.2000 \\ \hline
10       & 11     & 0.0004 & 0.0043 & 0.0729 \\ \hline
10       & 13     & 0.0004 & 0.0043 & 0.0729 \\ \hline
13       & 14     & 0.0009 & 0.0101 & 0.1723 \\ \hline
14       & 15     & 0.0018 & 0.0217 & 0.3660 \\ \hline
15       & 16     & 0.0009 & 0.0094 & 0.1710 \\ \hline
16       & 17     & 0.0007 & 0.0089 & 0.1342 \\ \hline
16       & 19     & 0.0016 & 0.0195 & 0.3040 \\ \hline
16       & 21     & 0.0008 & 0.0135 & 0.2548 \\ \hline
16       & 24     & 0.0003 & 0.0059 & 0.0680 \\ \hline
17       & 18     & 0.0007 & 0.0082 & 0.1319 \\ \hline
17       & 27     & 0.0013 & 0.0173 & 0.3216 \\ \hline
21       & 22     & 0.0008 & 0.0140 & 0.2565 \\ \hline
22       & 23     & 0.0006 & 0.0096 & 0.1846 \\ \hline
23       & 24     & 0.0022 & 0.0350 & 0.3610 \\ \hline
25       & 26     & 0.0032 & 0.0323 & 0.5130 \\ \hline
26       & 27     & 0.0014 & 0.0147 & 0.2396 \\ \hline
26       & 28     & 0.0043 & 0.0474 & 0.7802 \\ \hline
26       & 29     & 0.0057 & 0.0625 & 1.0290 \\ \hline
28       & 29     & 0.0014 & 0.0151 & 0.2490 \\ \hline
12       & 11     & 0.0016 & 0.0435 & 0.0000 \\ \hline
12       & 13     & 0.0016 & 0.0435 & 0.0000 \\ \hline
6        & 31     & 0.0000 & 0.0250 & 0.0000 \\ \hline
10       & 32     & 0.0000 & 0.0200 & 0.0000 \\ \hline
19       & 33     & 0.0007 & 0.0142 & 0.0000 \\ \hline
20       & 34     & 0.0009 & 0.0180 & 0.0000 \\ \hline
22       & 35     & 0.0000 & 0.0143 & 0.0000 \\ \hline
23       & 36     & 0.0005 & 0.0272 & 0.0000 \\ \hline
25       & 37     & 0.0006 & 0.0232 & 0.0000 \\ \hline
2        & 30     & 0.0000 & 0.0181 & 0.0000 \\ \hline
29       & 38     & 0.0008 & 0.0156 & 0.0000 \\ \hline
19       & 20     & 0.0007 & 0.0138 & 0.0000 \\ \hline
\end{tabular}
\end{table}
\clearpage

\begin{table}[]
\centering
\caption{Bus data. Data from Ref. \cite{IEEE39}.}
\label{table5}
\begin{tabular}{|c|c|c|c|c|c|c|c|c|}
\hline
Bus No. & $U$    & $D$ & $P_L^0$  & $P_L$    & $P_G^0$ & $P_G$  & Gen. No.        & $H$                 \\ \hline
1       & 1.0474 & 3   & 0.00     & 0.00     & 0.00    & 0.00   & -             & -                   \\ \hline
2       & 1.0487 & 3   & 0.00     & 0.00     & 0.00    & 0.00   & -             & -                   \\ \hline
3       & 1.0302 & 3   & -322.00  & -323.62  & 0.00    & 0.00   & -             & -                   \\ \hline
4       & 1.0039 & 3   & -500.00  & -501.62  & 0.00    & 0.00   & -             & -                   \\ \hline
5       & 1.0053 & 3   & 0.00     & 0.00     & 0.00    & 0.00   & -             & -                   \\ \hline
6       & 1.0077 & 3   & 0.00     & 0.00     & 0.00    & 0.00   & -             & -                   \\ \hline
7       & 0.9970 & 3   & -233.80  & -235.42  & 0.00    & 0.00   & -             & -                   \\ \hline
8       & 0.9960 & 3   & -522.00  & -523.62  & 0.00    & 0.00   & -             & -                   \\ \hline
9       & 1.0282 & 3   & 0.00     & 0.00     & 0.00    & 0.00   & -             & -                   \\ \hline
10      & 1.0172 & 3   & 0.00     & 0.00     & 0.00    & 0.00   & -             & -                   \\ \hline
11      & 1.0127 & 3   & 0.00     & 0.00     & 0.00    & 0.00   & -             & -                   \\ \hline
12      & 1.0002 & 3   & -7.50    & -9.12    & 0.00    & 0.00   & -             & -                   \\ \hline
13      & 1.0143 & 3   & 0.00     & 0.00     & 0.00    & 0.00   & -             & -                   \\ \hline
14      & 1.0117 & 3   & 0.00     & 0.00     & 0.00    & 0.00   & -             & -                   \\ \hline
15      & 1.0154 & 3   & -320.00  & -321.62  & 0.00    & 0.00   & -             & -                   \\ \hline
16      & 1.0318 & 3   & -329.00  & -330.62  & 0.00    & 0.00   & -             & -                   \\ \hline
17      & 1.0336 & 3   & 0.00     & 0.00     & 0.00    & 0.00   & -             & -                   \\ \hline
18      & 1.0309 & 3   & -158.00  & -159.62  & 0.00    & 0.00   & -             & -                   \\ \hline
19      & 1.0499 & 3   & 0.00     & 0.00     & 0.00    & 0.00   & -             & -                   \\ \hline
20      & 0.9912 & 3   & -628.00  & -629.62  & 0.00    & 0.00   & -             & -                   \\ \hline
21      & 1.0318 & 3   & -274.00  & -275.62  & 0.00    & 0.00   & -             & -                   \\ \hline
22      & 1.0498 & 3   & 0.00     & 0.00     & 0.00    & 0.00   & -             & -                   \\ \hline
23      & 1.0448 & 3   & -247.50  & -249.12  & 0.00    & 0.00   & -             & -                   \\ \hline
24      & 1.0373 & 3   & -308.60  & -310.22  & 0.00    & 0.00   & -             & -                   \\ \hline
25      & 1.0576 & 3   & -224.00  & -225.62  & 0.00    & 0.00   & -             & -                   \\ \hline
26      & 1.0521 & 3   & -139.00  & -140.62  & 0.00    & 0.00   & -             & -                   \\ \hline
27      & 1.0377 & 3   & -281.00  & -282.62  & 0.00    & 0.00   & -             & -                   \\ \hline
28      & 1.0501 & 3   & -206.00  & -207.62  & 0.00    & 0.00   & -             & -                   \\ \hline
29      & 1.0499 & 3   & -283.50  & -285.12  & 0.00    & 0.00   & -             & -                   \\ \hline
30      & 1.0475 & 2   & 0.00     & 0.00     & 250.00  & 248.38 & 10            & 500                 \\ \hline
31      & 0.9820 & 2   & -9.20    & -10.82   & 520.81  & 519.19 & 2             & 30.3                \\ \hline
32      & 0.9831 & 2   & 0.00     & 0.00     & 650.00  & 648.38 & 3             & 35.8              \\ \hline
33      & 0.9972 & 2   & 0.00     & 0.00     & 632.00  & 630.38 & 4             & 28.6              \\ \hline
34      & 1.0123 & 2   & 0.00     & 0.00     & 508.00  & 506.38 & 5             & 26.0              \\ \hline
35      & 1.0493 & 2   & 0.00     & 0.00     & 650.00  & 648.38 & 6             & 34.8              \\ \hline
36      & 1.0635 & 2   & 0.00     & 0.00     & 560.00  & 558.38 & 7             & 26.4              \\ \hline
37      & 1.0278 & 2   & 0.00     & 0.00     & 540.00  & 538.38 & 8             & 24.3              \\ \hline
38      & 1.0265 & 2   & 0.00     & 0.00     & 830.00  & 828.38 & 9             & 34.5              \\ \hline
39      & 1.0300 & 2   & -1104.00 & -1105.62 & 1000.00 & 998.38 & 1             & 42.0              \\ \hline
\end{tabular}
\end{table}

\clearpage
\bibliography{mybibphy}{}

\end{document}